\documentstyle[preprint,aps,epsfig,floats]{revtex}

\tightenlines
\draft
\begin{document}
\title{Perturbative Logarithms and Power Corrections in
       QCD Hadronic Functions.
       A Unifying Approach~\thanks{Invited plenary talk presented
       at the {\it 8th Adriatic Meeting on Particle Physics in the New
       Millennium}, Dubrovnik, Croatia, September 4-14, 2001.}
       }
\date{\today}
\preprint{{\hbox{RUB-TPII-01/02}}}
\author{N. G. Stefanis~\thanks{Email:
        stefanis@tp2.ruhr-uni-bochum.de
                               }
        }
\address{Institut f\"ur Theoretische Physik II \\
                  Ruhr-Universit\"at Bochum    \\
                  D-44780 Bochum, Germany      \\
           }

\maketitle        

\begin{abstract}
I present a unifying scheme for hadronic functions that comprises
logarithmic corrections due to gluon emission in perturbative QCD,
as well as power-behaved corrections of nonperturbative origin.
The latter are derived by demanding that perturbatively
resummed partonic observables should be analytic in the whole
$Q^2$-plane if they are to be related to physical observables
measured in experiments. I also show phenomenological consequences
of this approach. The focus is on the electromagnetic pion form
factor to illustrate both effects, Sudakov logarithms and power
corrections in leading order of $\Lambda_{\rm{QCD}}^2/Q^2$.
The same approach applied to the inclusive Drell-Yan cross section
enables us to perform an absolutely normalized calculation of the
leading power correction in $b^2\Lambda_{\rm{QCD}}^2$ (b being
the impact parameter), which after exponentiation, gives rise to a
nonperturbative Sudakov-type contribution that provides enhancement
rather than suppression, hence partly counteracting the
perturbative Sudakov suppression.
\end{abstract}
\pacs
\newpage
\input amssym.def
\input amssym.tex
\section{Introduction}
In recent years, effort in QCD has turned increasingly toward the
problem of including resummation effects due to multiple soft
gluon emission, both in perturbation theory, as well as in the
nonperturbative regime. The first effect is related to Sudakov
suppression~\cite{CS81}, well-known from QED, whereas those in the
nonperturbative regime manifest themselves as power-behaved
corrections~\cite{KS95}, which, after exponentiation, amount to a
Sudakov-like form factor~\cite{CSS85}. However, as it turns
out~\cite{KS01} this contribution provides enhancement rather than
suppression. The hope is that improving the perturbative and
nonperturbative structure of the theory this way, it will be
possible to get better agreement with the existing hadronic data
in terms of both correct overall shape and also normalization. In
these investigations the crucial organizing principle is QCD
factorization, which provides a handle to separate the
short-distance (hard) component of a reaction (controlled by the
large mass scale in the process, $Q$) - that will be treated
perturbatively - from its long-distance (soft) nonperturbative
part, related to the nontrivial QCD vacuum structure (and field
condensates).

In processes which involve the emission of virtual gluon quanta of
low momentum, one must resum their contributions to all orders of
the strong coupling constant. This gives rise to exponentially
suppressing factors in $b$-space (where $b$ is the impact parameter
conjugate to the transverse momentum $Q_{\perp}$) of the reaction
amplitude (or cross-section) of the Sudakov type with exponents
containing double and single logarithms of the large mass scale of
the process~\cite{CS81}.
However, because of the Landau singularity of the running coupling at
transverse distances $b\propto 1/\Lambda _{\rm{QCD}}$ , an essential
singularity appears in the Sudakov factor.
Thus, one has to consider power corrections of
${\cal O}\left( b^{2}\Lambda _{\rm{QCD}}^{2} \right)$, which, though
negligible for small $b$ relative to logarithmic corrections
$\propto\ln\left( b^{2}\Lambda _{\rm{QCD}}^{2} \right)$,
may become important for larger values of the impact parameter.

In this talk, I will discuss a general methodology to treat (power)
series in the running strong coupling in connection with gluon
emission. To be more precise, I will address this issue in terms
of two processes: one to which the OPE applies, viz. the pion
electromagnetic form factor at leading perturbative order, and
another, the Drell-Yan process, to which the OPE is not applicable.
The first is a typical example of an exclusive process with
registered intact hadrons in the initial and final states (for a
recent review and references, see, e.g.,~\cite{Ste99}).
Such processes provide a ``window'' to view the detailed structure
of hadrons in terms of quarks and gluons at Fermi level
({\it Hadron Femptoscopy}).
The Drell-Yan mechanism, on the other hand, has two identified
hadrons in the initial state and a lepton pair (plus unspecified
particles) in the final state, whose transverse momentum
distribution is proportional to the large invariant mass
of the materialized photon.

The goal in the second case will be to obtain not only the usual
resummed (Sudakov) expression (which comprises logarithmic
corrections due to soft-gluon radiation), but also to include the
leading power correction as well, specifying, in particular, its
concomitant coefficient.
This  becomes possible within a theoretical scheme, which models the IR
behavior of the running coupling by demanding analyticity of physical
observables (in the complex $Q^{2}$ plane) as a {\it whole} -- as
opposed to imposing analyticity of individual powers, i.e., order by
order in perturbation theory --, while preserving renormalization-group
invariance (references and additional information can be found in the
recent surveys~\cite{Shi01,SS99} and D.V. Shirkov, these proceedings).
The underlying idea behind our method~\cite{KS01}, is to demand
that if hadronic observables, calculated at the partonic level,
are to be compared with experimental data, they have to be
analytic in the entire $Q^2$ plane. This ``analytization''
procedure encompasses Renormalization Group (RG) invariance (i.e.,
resummation of UV logarithms and correct UV asymptotics) and
causality (which imposes a spectral representation). As we shall
see below, {\it analytization} removes all unphysical singularities
in the the IR region, rendering perturbatively calculated hadronic
observables IR-renormalon free.

\section{Analytic Factorization Scheme (AFS)}
\subsection*{2.1 Perturbative Pion Form factor with Sudakov Corrections}
Let us conduct our investigation by considering the space-like
electromagnetic pion's form factor in the transverse (impact)
configuration space:
\begin{eqnarray}
\everymath{\displaystyle}
  F_{\pi}\left(Q^{2}\right)
= &&
  \int_{0}^{1} dx dy \int_{-\infty}^{\infty}
                     \frac{d^{2}\bf{b}}{(4\pi )^{2}} \,
          {\cal P}_{\pi}^{\rm out}
          \left( y, b, P^{\prime}; C_{1}, C_{2}, C_{4}
          \right)
  T_{\rm H}\left(
                 x, y, b, Q; C_{3}, C_{4}
           \right)
\nonumber \\
&&
\times\;  {\cal P}_{\pi}^{\rm in}
                   \left( x, b, P; C_{1}, C_{2}, C_{4}
                   \right) + \ldots ,
\label{eq:piffbspace}
\end{eqnarray}
where the modified pion wave function is defined in terms of matrix
elements, viz.,
\begin{eqnarray}
  {\cal P}_{\pi}
                \left( x, b, P, \mu
                \right)
& = &
  \int_{}^{|\bf{k}_{\perp} |<\mu} d^{2}\bf{k}_{\perp}
  {\rm e}^{- i \bf{k}_{\perp} \cdot \bf{b}_{\perp}}
  {\tilde{\cal P}}_{\pi}\left( x, \bf{k}_{\perp} , P \right)
\nonumber \\
& = &
  \int_{}^{}\!\! \frac{dz^{-}}{2\pi}
  {\rm e}^{ -ix P^{+}z^{-}}\!
  {\left\langle 0 \left\vert
  {\rm T} \left(
                \bar{q}(0)\gamma ^{+}\gamma _{5}
                q\left(0,z^{-},\bf{b}_{\perp} \right)
          \right)
  \right\vert \pi (P) \right\rangle}_{A^{+}=0}
\label{eq:matrel}
\end{eqnarray}
with $P^{+}=Q/\sqrt{2}=P^{-\prime}$, $Q^{2}=-(P^{\prime}-P)^{2}$,
whereas the dependence on the renormalization scale $\mu$ on the
RHS of (\ref{eq:matrel}) enters through the normalization scale of the
current operator, evaluated on the light cone, and the dependence on
the effective quark mass has not been displayed explicitly.
In (\ref{eq:matrel}), $T_{\rm H}$ is the amplitude for a quark
and an anti-quark to scatter via a series of hard-gluon exchanges with
gluonic transverse momenta (alias inter-quark transverse distances) not
neglected from the outset.
In the above, the ellipsis indicates the non-factorizing soft
part, as well as disregarded higher-order corrections.
The scheme constants $C_{i}$ emerge from the truncation of the
perturbative series and would be absent if one was able to derive
all-order expressions in the coupling constant.
The scale $C_{1}/b$ ($C_{1}=C_{3}$) serves to separate perturbative
from non-perturbative transverse distances (lower factorization scale
of the Sudakov regime and {\it transverse} cutoff).
The re-summation range in the Sudakov form factor is limited from
above by the scale $C_{2}\xi Q$ (upper factorization scale of the
Sudakov regime and {\it collinear} cutoff).\footnote{Note that
$\sqrt{2}C_{2}= C_{2}^{{\rm CSS}}$~\cite{CS81}.}
The arbitrary constant $C_{4}$ serves to define the renormalization
scale $C_{4}f(x,y)Q = \mu _{\rm R}$, which appears in the argument
of the analytic running coupling
$\alpha _{\rm s}^{\rm{an}}$~\cite{SS97} (choice of
renormalization prescription):
\begin{eqnarray}
    \bar{\alpha}_{\rm s}^{\rm{an}(1)}(Q^{2})
& \equiv &
    \bar{\alpha}_{\rm s}^{\rm{pert}(1)}(Q^{2})
  + \bar{\alpha}_{\rm s}^{\rm{npert}(1)}(Q^{2})
\nonumber \\
& = &
  \frac{4\pi}{\beta _{0}}
  \left[
          \frac{1}{\ln \left( Q^{2}/\Lambda ^{2} \right)}
        + \frac{\Lambda ^{2}}{\Lambda ^{2} - Q^{2}}
  \right],
\label{eq:oneloopalpha_an}
\end{eqnarray}
where here and below $\Lambda \equiv \Lambda _{\rm{QCD}}$ is the
QCD scale parameter.

To leading order in analytic perturbation theory (APT), one has
\begin{equation}
  T_{\rm H}\left(
                 x, y, b, Q; \mu _{\rm R}
           \right)
=
  8 C_{\rm F} \alpha _{\rm s}^{\rm{an}}(\mu _{\rm R}^{2})
  K_{0} \left( \sqrt{xy}\, bQ\right),
\label{eq:T_Hbspace}
\end{equation}
where
$C_{\rm F}=(N_{\rm c}^{2} - 1)/2N_{\rm c} = 4/3$ for
$SU(3)_{\rm c}$.
The amplitude
\begin{eqnarray}
\everymath{\displaystyle}
  {\cal P}_{\pi}\left( x, b, P\simeq Q, C_{1}, C_{2}, \mu
                \right)
=
 \exp \Biggl[
             && -  s\left( x, b, Q, C_{1}, C_{2} \right)
                -  s\left( \bar{x}, b, Q, C_{1}, C_{2} \right)
\nonumber \\
&&
                - 2 \int_{C_{1}/b}^{\mu} \!\frac{d\bar\mu}{\bar\mu} \,
                \gamma _{\rm q}\left(\alpha _{\rm s}^{\rm{an}}(\bar\mu )
                           \right)
      \Biggr]
  {\cal P}_{\pi}\left( x, b, C_{1}/b \right)
\label{eq:piamplbspace}
\end{eqnarray}
describes the distribution of longitudinal momentum fractions of the
q\=q pair, taking into account the intrinsic transverse size of the
pion state and comprising corrections due to soft real and
virtual gluons, including also evolution from the initial
amplitude ${\cal P}_{\pi}\left( x, b, C_{1}/b \right)$ at scale
$C_{1}/b$ to the renormalization scale $\mu \propto Q$
(more details and references are relegated to~\cite{SSK00}).
The main effect of the absence of a Landau pole in the running coupling
$\alpha _{\rm s}^{\rm{an}}$ is to make the functions
$s\left( x, b, Q, C_{1}, C_{2} \right)$,
$s\left( \bar{x}, b, Q, C_{1}, C_{2} \right)$
well-defined (analytic) in the IR region and to slow down evolution by
extending soft-gluon cancellation down to the scale
$C_{1}/b \simeq \Lambda _{\rm{QCD}}$, where the full Sudakov form
factor acquires a finite value, modulo its $Q^{2}$ dependence (see
LHS of Fig.~\ref{fig:ste_eps1}).
In addition, as we shall see below, the Sudakov exponent contains
power-behaved corrections in
$\left(C_{1}/b\Lambda\right)^{2p}$ and
$\left(C_{2}/\xi Q\Lambda\right)^{2p}$, starting with $p=1$.
Such contributions are the footprints of soft gluon emission at the
kinematic boundaries to the non-perturbative QCD regime, characterized
by the transversal (or IR) and the longitudinal (or collinear) cutoffs.

\input psbox.tex
\begin{figure}
\begin{picture}(0,40)
  \put(60,-60){\psboxscaled{400}{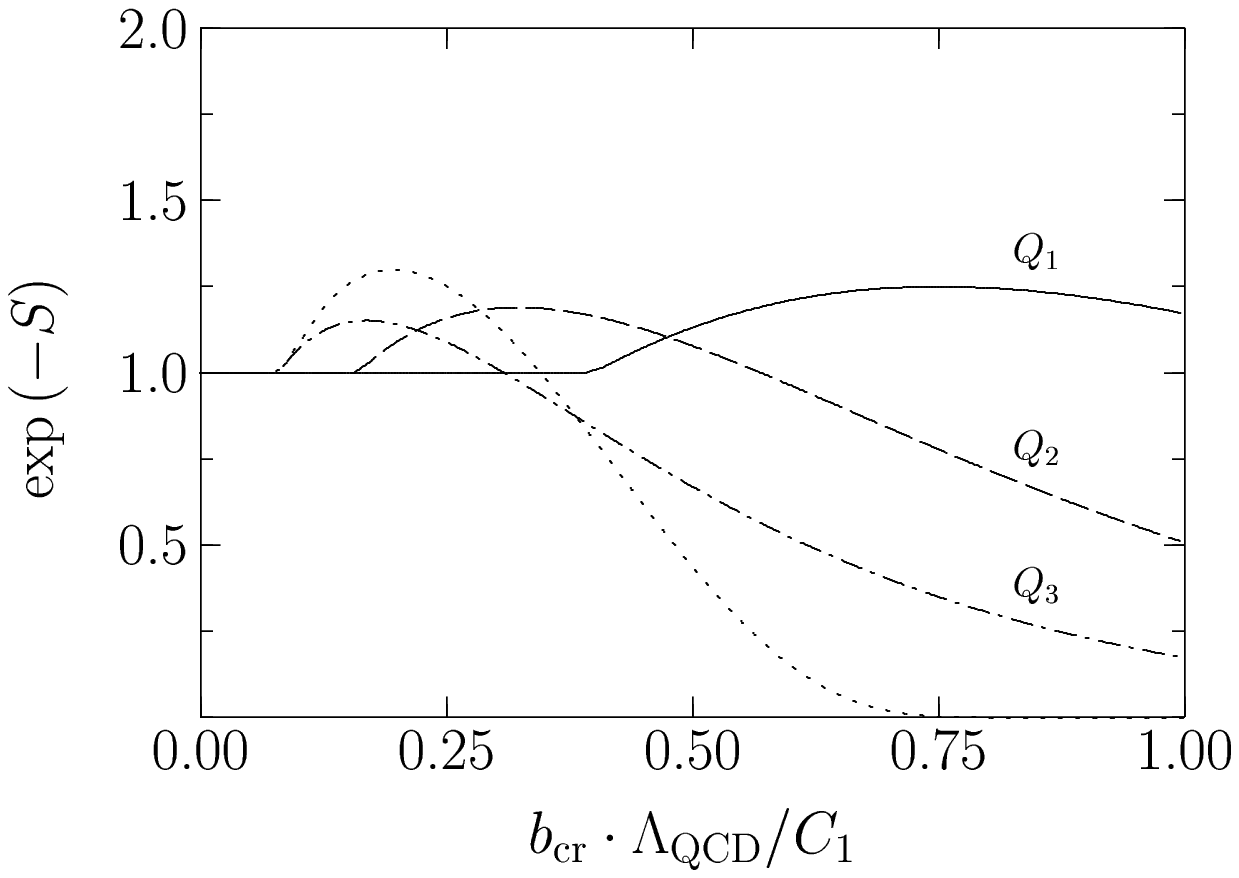}}
\end{picture}
\begin{picture}(0,40)
  \put(240,-60){\psboxscaled{450}{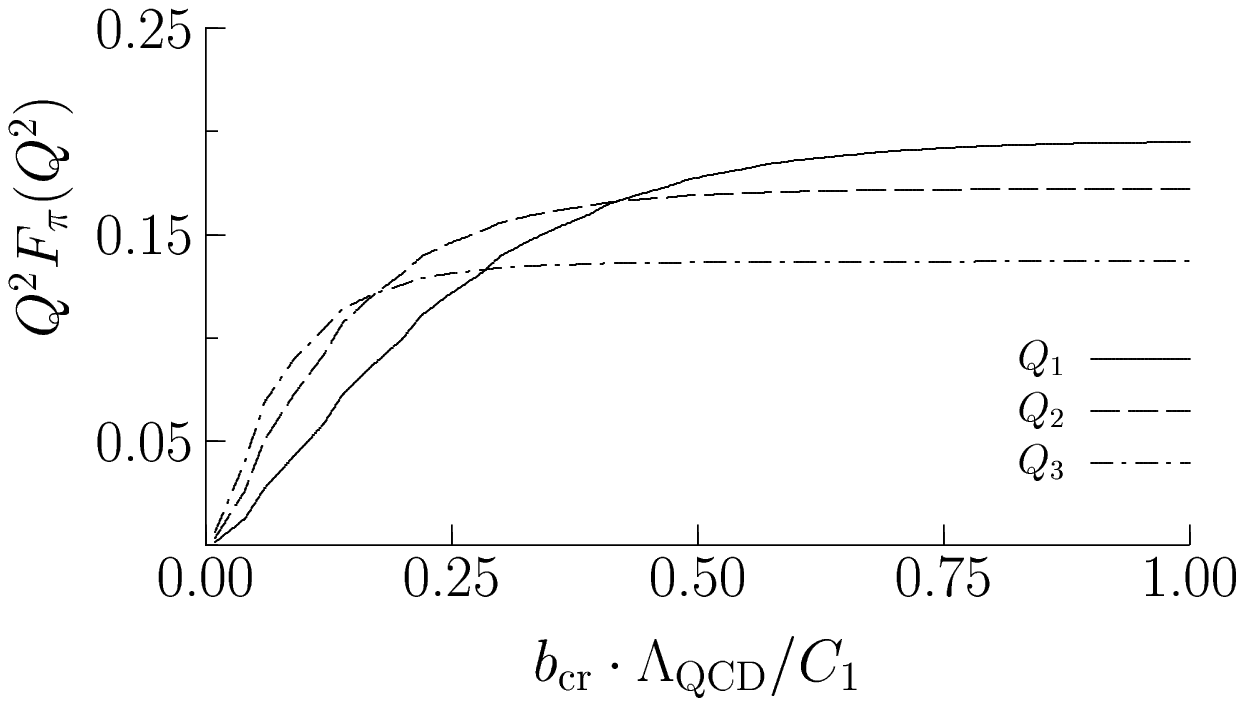}}
\end{picture}
\vspace{2.5 true cm}
\caption[]{(a) Sudakov form factor versus transverse separation
           $b$ for three $Q^{2}$ values:
           $Q_{1}=2$~GeV, $Q_{2}=5$~GeV, and $Q_{3}=10$~GeV, with all
           $\xi _{i}=1/2$, and where we have set
           $C_{1}=2{\rm e}^{-\gamma _{\rm E}}$,
           $C_{2}={\rm e}^{-1/2}$ and
           $\Lambda _{\rm{QCD}}=0.242~{\rm{GeV}}$.
           The dotted curve shows the result obtained with
           $\alpha _{\rm s}^{\overline{\rm{MS}}}$, and
           $\Lambda _{\rm{QCD}}=0.2~{\rm{GeV}}$ for $Q_{2}=5$~GeV, using
           the same set of $C_{i}$.
           In that case, evolution is limited by the (renormalization) scale
           $\mu_{\rm R}=t=\{ {\rm{max}}\sqrt{xy}\, Q, C_{1}/b \}$,
           as proposed in~\cite{LS92}, albeit the enhancement at small
           $b$-values due to the quark anomalous dimension is not neglected
           here.
           (b) Saturation behavior of pion's electromagnetic
           form factor, calculated in the AFS at NLO with commensurate scale
           setting (see text) and including a mass term (with
           $m_{\rm q}=0.33$~GeV) in the BHL ansatz~\cite{BHL83} for the
           soft pion wave function. The scheme parameters are defined
           in (\ref{eq:Cs}). Here $b_{\rm{cr}}$ denotes the
           integration cutoff over transverse distances in (\ref{eq:NLOpiff}).
           The momentum transfer values are as in part (a)}
\label{fig:ste_eps1}
\end{figure}

The pion distribution amplitude evaluated at the (low) factorization
scale $C_{1}/b$ is approximately given by
\begin{equation}
  {\cal P}_{\pi}\left( x, b, C_{1}/b, m_{\rm q} \right)
\simeq
  \frac{f_{\pi}/\sqrt{2}}{2\sqrt{N_{\rm c}}} \,
  \phi _{\pi}\left( x, C_{1}/b \right)
  \Sigma \left( x, b, m_{\rm q} \right).
\label{eq:inpiampl}
\end{equation}

To model the intrinsic transverse momenta of the pion bound
state, we have to make an ansatz for their distribution.
(For a recent derivation from an instanton-based model,
see~\cite{PR01}). Here, I employ the Brodsky-Huang-Lepage (BHL)
ansatz~\cite{BHL83} and parameterize the distribution
$\Sigma (x, \bf{k}_{\perp}, m_{\rm q})$
in the intrinsic transverse momentum $k_{\perp}$
(or equivalently the intrinsic inter-quark transverse distance $b$)
in the form of a {\it non-factorizing} in the variables $x$ and
$k_{\perp}$ (or $x$ and $b$) Gaussian function:
\begin{equation}
  \Psi _{\pi} \left(x, \bf{k}_{\perp}, C_{1}/b, m_{\rm q}\right)
=
  \frac{f_{\pi}/\sqrt{2}}{2N_{\rm c}}
  \Phi _{\pi}(x, C_{1}/b)
  \Sigma \left(x, \bf{k}_{\perp}, m_{\rm q}\right),
\label{eq:psiansatz}
\end{equation}
where
\begin{equation}
  \Phi _{\pi}\left( x, C_{1}/b \right)
=
  A\, \Phi _{\rm{as}}(x)
=
  A\, 6x(1-x)
\label{eq:asymptoticwf}
\end{equation}
is the asymptotic distribution amplitude, with $A$ being an appropriate
normalization factor, and where
\begin{equation}
  \Sigma \left(x, \bf{k}_{\perp}, m_{\rm q}\right)
=
  16 \pi ^{2} \beta^{2}_{\pi} g(x)
  \hat\Sigma \left(x, \bf{k}_{\perp}\right)
  \hat\Sigma \left(x, m_{\rm q}\right)
\label{eq:FullGaussiankspace}
\end{equation}
with
\begin{equation}
  \hat\Sigma \left(x, \bf{k}_{\perp}\right)
=
  \exp{\left[-\beta^{2}_{\pi} \bf{k}_{\perp}^{2} g(x)\right]},
\label{eq:k_perpGaussian}
\end{equation}
and
\begin{equation}
  \hat\Sigma \left(x, m_{\rm q}\right)
=
  \exp{\left[-\beta^{2}_{\pi} m_{\rm q}^{2} g(x)\right]}.
\label{eq:massGaussian}
\end{equation}
By inputting $f_{\pi}$ and the value of the quark mass $m_{\rm q}$
and using $g(x)=1/\left(x\bar{x}\right)$, with $\bar{x}\equiv (1-x)$,
we determine the parameters (we refer for more details
to~\cite{SSK00})
$A$, $\beta^{2}_{\pi}$, $P_{\rm{q\bar q}}$, and
${\langle \bf{k}_{\perp}^{2}\rangle}^{1/2}$, tabulated in
Table
~\ref{tab:parameters}.

\begin{table}
\caption[av]{Values of parameters entering the pion wave
         function~\cite{SSK00}.
         The values in parentheses refer to the case $m_{\rm q}=0$
         and the subscript ``as'' on $\beta^{2}_{\pi}$ to the asymptotic
         distribution amplitude}
\begin{center}
\renewcommand{\arraystretch}{1.4}
\setlength\tabcolsep{5pt}
\begin{tabular}{lc}
\noalign{\smallskip}
Input parameters & Determined parameters \\
\noalign{\smallskip}
$m_{\rm q}=0.33$~GeV & $A=\frac{1}{6}\cdot 10.01$
$(\frac{1}{6}\cdot 6)$\\
$f_{\pi}=0.1307$~GeV & $\beta_{\rm{as}}^{2}=0.871$~GeV${}^{-2}$
$(0.743$~GeV${}^{-2})$\\
& ${\langle \bf{k}^{2}\rangle}^{1/2}=0.352$~GeV
$(0.367$~GeV)\\
& $P_{\rm{q\bar q}}=0.306$
$(0.250)$\\
\end{tabular}
\end{center}
\label{tab:parameters}
\end{table}

We have now to calculate the Sudakov contribution within the AFS.
Generically, the Sudakov form factor
$F_{\rm S}\left(\xi , b, Q, C_{1}, C_{2}\right)$,
i.e., the exponential factor in front of the wave function, will be
expressed as the expectation value of an open Wilson (world) line along
a contour of finite extent, $C$, which follows the bent quark line in
the hard-scattering process from the segment with direction
(four-momentum) $P$ to that with direction $P^{\prime}$ after being
abruptly derailed by the hard interaction which creates a ``cusp'' in
$C$. It is to be evaluated within the range of momenta from $C_{1}/b$
(IR cutoff) to $C_{2}\xi Q$ (longitudinal cutoff) (where
$\xi = x, \bar{x}, y, \bar{y}$) and the region of hard interaction of
the Wilson line with the off-shell photon is factorized out.
Then the Sudakov functions, entering (\ref{eq:piamplbspace}), can be
expressed in terms of the momentum-dependent cusp anomalous dimension
of the bent contour~\cite{CS81,KR87,Kor89,GKKS97} to read
\begin{equation}
  s\left(\xi , b, Q, C_{1}, C_{2} \right)
=
  \frac{1}{2}
  \int_{C_{1}/b}^{C_{2}\xi Q}
  \frac{d\mu}{\mu} \,
  \Gamma _{\rm{cusp}}
        \left(\gamma , \alpha _{\rm s}^{\rm{an}}(\mu )
        \right)
\label{eq:sudfuncusp}
\end{equation}
with the anomalous dimension of the cusp given by
\begin{eqnarray}
  \Gamma _{\rm{cusp}}
        \left( \gamma , \alpha _{\rm s}^{\rm{an}}(\mu ) \right)
& = &
  2 \ln \left(
              \frac{C_{2}\xi Q}{\mu}
        \right) A\left( \alpha _{\rm s}^{\rm{an}}(\mu ) \right)
              + B\left( \alpha _{\rm s}^{\rm{an}}(\mu ) \right) \; ,
\nonumber \\
& \equiv &
    \Gamma _{\rm{cusp}}^{\rm{pert}}
  +
    \Gamma _{\rm{cusp}}^{\rm{npert}},
\label{eq:gammacusp}
\end{eqnarray}
$\gamma = \ln \left(C_{2}\xi Q/\mu \right)$
being the cusp angle, i.e., the emission angle of a soft gluon and the
bent eikonalized quark line after the external (large) momentum $Q$ has
been injected at the cusp point by the off-mass-shell photon, and where
in the second line of (\ref{eq:gammacusp}) the superscripts relate
to the origin of the corresponding terms in the running coupling.
The functions $A$ and $B$ are known at two-loop order:
\begin{eqnarray}
\everymath{\displaystyle}
  A\left( \alpha _{\rm s}^{\rm{an}}(\mu ) \right)
& = &
  \frac{1}{2}
  \left[
         \gamma _{\cal K}
                 \left( \alpha _{\rm s}^{\rm{an}}(\mu ) \right)
       + \beta (g) \frac{\partial}{\partial g}
       {\cal K}(C_{1}, \alpha _{\rm s}^{\rm{an}}(\mu ))
  \right]
\nonumber \\
& = &
       C_{\rm F} \frac{\alpha _{\rm s}^{\rm{an}}(g(\mu ))}{\pi}
     + \frac{1}{2} K\left( C_{1} \right) C_{\rm F}
       \left( \frac{\alpha _{\rm s}^{\rm{an}}(g(\mu ))}{\pi}
       \right)^{2},
\label{eq:funAnlo}
\end{eqnarray}
and
\begin{eqnarray}
\everymath{\displaystyle}
  B\left( \alpha _{\rm s}^{\rm{an}}(\mu ) \right)
& = &
  - \frac{1}{2}
    \left[
          {\cal K} \left(
                         C_{1}, \alpha _{\rm s}^{\rm{an}}(\mu )
                   \right)
        + {\cal G} \left(
                         \xi, C_{2}, \alpha_{\rm s}^{\rm{an}}(\mu )
                   \right)
    \right]
\nonumber \\
& = &
      \frac{2}{3} \frac{\alpha _{\rm s}^{\rm{an}}(g(\mu ))}{\pi}
      \ln \left(
          \frac{C_{1}^{2}}{C_{2}^{2}}\frac{{\rm e}^{2\gamma_{\rm E}-1}}{4}
          \right).
\label{eq:funBnlo}
\end{eqnarray}
The first term in (\ref{eq:funAnlo}) is universal,\footnote{In
works quoted above, the cusp anomalous dimension is identified with the
universal term, whereas the other (scheme and/or process dependent)
terms are considered as additional anomalous dimensions.
Here this distinction is irrelevant.}
while the second one as well as the contribution termed $B$ are scheme
dependent.
The K-factor in the ${\overline{\rm MS}}$ scheme to two-loop order
is given by \cite{CS81,KR87,KT82}
\begin{equation}
  K\left(C_{1}\right)
=
     \left(\frac{67}{18} - \frac{\pi ^{2}}{6}\right) C_{\rm A}
   - \frac{10}{9}n_{\rm f} T_{\rm F}
   + \beta _{0} \ln \left( C_{1} {\rm e}^{\gamma _{\rm E}}/2 \right)
\label{eq:Kfactor}
\end{equation}
with
$C_{\rm A}=N_{\rm C}=3$,
$n_{\rm f}=3$,
$T_{\rm F}=1/2$,
and $\gamma _{\rm E}$ being the Euler-Mascheroni constant.
A set of constants \hbox{$C_{i}$, $(i=1,2,3)$}, which eliminate
artifacts of dimensional regularization while practically preserving
the matching between the re-summed and the fixed-order calculation,
are~\cite{SSK00}
\begin{eqnarray}
  C_{1}
& = &
2\exp \left(-\gamma _{\rm{E}}\right),
\;\;\;
  C_{2} = \exp \left(-1/2\right),
\;\;\;
  C_{3} = 2\exp \left(-\gamma _{\rm{E}}\right),
\;\;\;
  C_{4} = \exp \left(-4/3\right),
\nonumber \\
  K
& = &
  4.565,
\;\;\;
  \kappa = 0.
\label{eq:Cs}
\end{eqnarray}

The quantities ${\cal K}$, ${\cal G}$ in (\ref{eq:funBnlo}) are
calculable using the non-Abelian extension to QCD~\cite{CS81} of
the Grammer-Yennie method for QED or employing the Wilson (world)
lines approach~\cite{KR87,Kor89,GKKS97}.
The soft (Sudakov-type) form factor depends only on the cusp angle
which varies with the inter-quark transverse distance $b$ ranging
between $C_{1}/b$ and $C_{2}\xi Q$. The corresponding anomalous
dimensions are inter-linked through the relation
$
  2 \Gamma _{\rm{cusp}}\left(\alpha_{\rm s}^{\rm{an}}(\mu )
                           \right)
=
  \gamma _{\cal K}\left(\alpha_{\rm s}^{\rm{an}}(\mu )\right)
$
with
$
 \Gamma _{\rm{cusp}}(\alpha_{\rm s}^{\rm{an}}(\mu ))
=
  C_{\rm F}\, \alpha_{\rm s}^{\rm{an}}(\mu ^{2})/\pi
$, which shows that
$
 \frac{1}{2}\gamma_{\cal K}
=
 A\left( \alpha_{\rm s}^{\rm{an}}(\mu ) \right)
$.
(Note that $\gamma _{\cal G} = - \gamma_{\cal K}$ and $
 \gamma_{\rm q}\left( \alpha_{\rm s}^{\rm{an}}(\mu )
                   \right)
=
 - \alpha_{\rm s}^{\rm{an}}(\mu ^{2})/\pi
$.)

The leading contribution to the Sudakov functions
$s\left( \xi ,b, Q, C_{1}, C_{2} \right)$
(where $\xi = x, \bar{x}, y, \bar{y}$) within our framework,
is obtained by expanding the functions $A$ and $B$ in a power
series in $\alpha _{\rm s}^{\rm{an}}$ and collecting
together all large logarithms
$
 \left( \frac{\alpha _{\rm s}^{\rm{an}}}{\pi} \right)^{n}
 \ln \left ( \frac{C_{2}}{C_{1}} \xi b Q \right)^{m}
$,
which correspond to large logarithms
$
 \ln \left( \frac{Q^{2}}{k_{\perp}^{2}} \right)
$
in transverse momentum space.
The leading contribution results from the expression
\begin{eqnarray}
\everymath{\displaystyle}
  s\left( \xi , b, Q, C_{1}, C_{2}\right)
= &&
    \frac{1}{2}
    \int_{C_{1}/b}^{C_{2}\xi Q} \frac{d\mu}{\mu}
    \Biggl\{
           2 \ln \left( \frac{C_{2} \xi Q}{\mu} \right)
       \Biggl[
         \frac{\alpha _{\rm s}^{{\rm an}(2)}(\mu )}{\pi}
         A^{(1)}
\nonumber \\
&&
\!\!\!\!\!\!\!\!\!\!\!\!\!\!\!\!\!\!\!\!\!\!\!\!\!\!\!\!\!\!\!\!\!\!\!
       +{\left(
         \frac{\alpha _{\rm s}^{\rm{an}(1)}(\mu )}{\pi}
         \right)}^{2}
         A^{(2)}\left( C_{1} \right)
       \Biggr]
      +  \frac{\alpha _{\rm s}^{{\rm an}(1)}(\mu )}{\pi}
         B^{(1)}\left( C_{1}, C_{2} \right)
      +  \ldots
    \Biggr\},
\label{eq:SudakovNLO}
\end{eqnarray}
where the two-loop expression~\cite{SS97} for the strong coupling
is to be used in front of $A^{(1)}$, whereas the other two terms
are to be evaluated with the one-loop result.
Let me remark at this point that in the following we ignore the
difference between the analytic strong coupling squared and its
``analytized'' second power. These issues will be considered
elsewhere.
The specific values of the coefficients $A^{(i)}$, $B^{(i)}$ are
\begin{eqnarray}
\everymath{displaystyle}
  A^{(1)}                         & = & C_{\rm F} \; ,
\nonumber \\
  A^{(2)} \left(C_{1}\right)      & = & \frac{1}{2}\, C_{\rm F} \,
                                        K\left( C_{1} \right) \; ,
\nonumber \\
  B^{(1)}\left(C_{1},C_{2}\right) & = & \frac{2}{3}
                \ln \left(
                          \frac{C_{1}^{2}}{C_{2}^{2}}
                          \frac{{\rm e}^{2\gamma_{\rm E}-1}}{4}
                    \right),
\label{eq:expcoef}
\end{eqnarray}
in which the term proportional to $A^{(1)}$ represents the universal
part.
The {\it universal} part of the Sudakov factor in LLA and including
power corrections, reads
\begin{eqnarray}
\everymath{\displaystyle}
  F_{\rm S}^{\rm{univ}}\left(\mu _{\rm F}, Q\right)
=
&&
  \exp \Biggl\{
               - \frac{C_{\rm F}}{\beta _{0}}
       \Biggl[
                 \ln \left(\frac{\tilde{Q}^{2}}{\Lambda ^{2}}\right)
                 \ln \frac{\ln \tilde{Q}^{2}/\Lambda ^{2}}
                          {\ln \mu _{\rm F}^{2}/\Lambda ^{2}}
               - \ln\frac{\tilde{Q}^{2}}{\mu _{\rm F}^{2}}
  + \ln\left(\frac{\tilde{Q}^{2}}{\mu _{\rm F}^{2}}\right)
\nonumber \\
&&  \times
    \ln\frac{{\Lambda}^{2}-\mu _{\rm F}^{2}}{\Lambda ^{2}}
  + \frac{1}{2}\, \ln ^{2} \frac{\tilde{Q}^{2}}{\mu _{\rm F}^{2}}
  + {\rm Li}_{2}\left(\frac{\tilde{Q}^{2}}{\Lambda ^{2}}\right)
  \! - {\rm Li}_{2}\left(\frac{\mu _{\rm F}^{2}}{\Lambda ^{2}}\right)
       \Biggr]\!
       \Biggl\},
\label{eq:unipart}
\end{eqnarray}
where $\tilde{Q}$ represents the scale $C_{2}\xi Q$ and the IR matching
(factorization) scale $\mu _{\rm F}$ varies with the inverse
transverse distance $b$, i.e., $\mu _{\rm F} = C_{1}/b$.
Note that the four last terms in this equation originate from the
non-perturbative power correction (cf.~(\ref{eq:gammacusp})), and
that ${\rm Li}_{2}$ is the dilogarithm (Spence) function which
comprises power-behaved corrections of the IR-cutoff ($b\Lambda$)
and the longitudinal cutoff ($Q/\Lambda$).
To complete the discussion about the Sudakov factor, I display the
result obtained by neglecting power corrections:
\begin{eqnarray}
\everymath{\displaystyle}
  s\left( \xi , b, Q, C_{1}, C_{2} \right)
= &&
  \frac{1}{\beta _{0}}
  \left[
         \left(
               2 A^{(1)} \hat Q + B^{(1)}
         \right)\ln \frac{\hat Q}{\hat b}
       - 2 A^{(1)} \left( \hat Q - \hat b \right)
  \right]
 - \frac{4}{\beta _{0}^{2}}
     A^{(2)}
\nonumber \\
&& \times
     \left(
           \ln \frac{\hat Q}{\hat b} -
           \frac{\hat{Q} - \hat{b}}{\hat b}
     \right)
  + \frac{\beta _{1}}{\beta _{0}^{3}} A^{(1)}
     \Biggl\{
              \ln \frac{\hat Q}{\hat b}
            - \frac{\hat Q - \hat b}{\hat b}
              \left[
                    1 + \ln \left( 2 \hat b \right)
              \right]
\nonumber \\
       &&      + \frac{1}{2}
              \left[
                     \ln ^{2}\left(2\hat Q \right)
                   - \ln ^{2}\left(2\hat b \right)
              \right]
     \Biggr\},
\label{eq:oldsudakov}
\end{eqnarray}
where the convenient abbreviations \cite{LS92}
$
 \hat Q
\equiv
 \ln \frac{C_{2} \xi Q}{\Lambda}
$
and
$
 \hat b
\equiv
  \ln \frac{C_{1}}{b\Lambda}
$
have been used. Note that expressions given in the literature
by other authors are erroneous.

\begin{figure}
\tighten
\begin{center}
\centerline{\epsfig{file=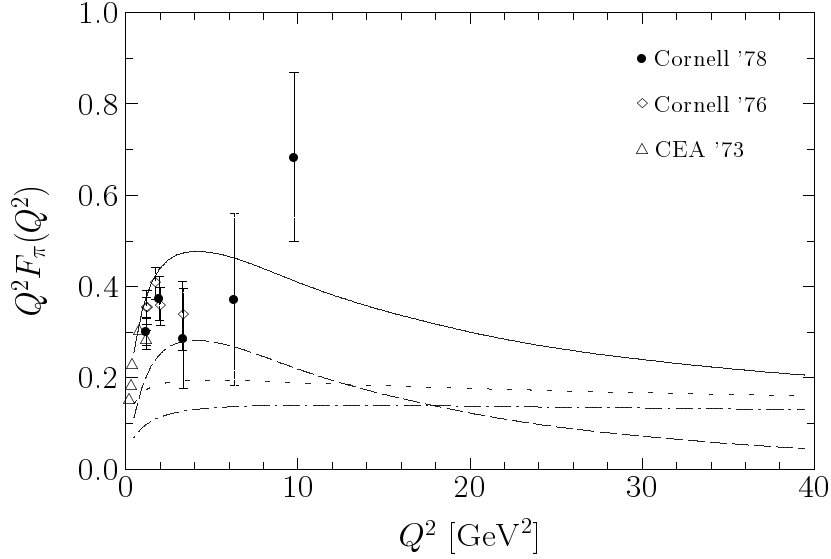,height=8cm,width=12cm,silent=}}
\end{center}
\vspace{1cm}
\caption[]{Space-like pion form factor calculated within the AFS.
         Further details are provided in the text.
         LO calculation (dashed-dotted line); NLO calculation
         (dotted line).
         The dashed line gives the result for the soft, Feynman-type
         contribution, computed with $m_{\rm q}=0.33$~GeV in the pion
         wave function, and the solid line represents the sum of the
         NLO hard contribution and the soft one~\cite{SSK00}.
         The data are taken from~\cite{Bro73,Beb76}.}
\label{fig:ste_eps2}
\end{figure}

In the following, (\ref{eq:SudakovNLO}) is evaluated
numerically to NLLA with appropriate kinematic bounds~\cite{SSK00}
to ensure proper factorization at the numerical level.
The electromagnetic pion form factor in next-to-leading logarithmic
order has the following form in LO of $T_{\rm H}$:
\begin{eqnarray}
\everymath{\displaystyle}
  F_{\pi}(Q^{2})
= &&
  \frac{2}{3}\, A^{2} \pi\, C_{\rm F}\, f_{\pi}^{2}\,
  \int_{0}^{1} dx
  \int_{0}^{1} dy
  \int_{0}^{\infty}b\, db\,
  \alpha _{\rm s}^{{\rm{an}}(1)}\left(\mu _{\rm R}\right)
  \Phi _{\pi} (x)
  \Phi _{\pi} (y)
\nonumber \\
&& \times
  \exp \left[-\frac{b^{2} \left(x\bar{x} + y\bar{y}\right)}
  {4\beta_{\pi}^{2}}\right]
  \exp{ \left[-\beta_{\pi}^{2} m_{\rm q}^{2}
  \left(\frac{1}{x\bar x} + \frac{1}{y\bar y}\right)
        \right]}
  K_{0}\left(\sqrt{xy}Qb\right)
\nonumber \\
&& \times
   \exp \left[ - S \left(x,y,b,Q, C_{1},C_{2},C_{4}\right) \right],
\label{eq:pifofafin}
\end{eqnarray}
whereas in NLO it reads
\begin{eqnarray}
\everymath{displaystyle}
  F_{\pi}\left( Q^{2} \right)
& = &
  16 A^{2} \pi C_{\rm F}
  \left(
        \frac{f_{\pi}/\sqrt{2}}{2\sqrt{N_{\rm c}}}
  \right)^{2}
  \int_{0}^{1} dx \int_{0}^{1} dy
  \int_{0}^{\infty} b \,db \,
  \alpha_{\rm s}^{\rm{an}}\left( \mu_{\rm R}^{2}\right)
  \Phi _{\pi}(x) \Phi _{\pi}(y)
\nonumber \\
&& \times
  \exp\left[
            - \frac{b^{2}\left( x\bar x + y\bar y \right)}
              {4\beta_{\pi}^{2}}
      \right]
  \exp{ \left[-\beta_{\pi}^{2} m_{\rm q}^{2}
  \left(\frac{1}{x\bar x} + \frac{1}{y\bar y}\right)
        \right]}
  K\left(\sqrt{x y}Qb\right)
\nonumber \\
&& \times
  \exp\left( - S\left(x, y, b, Q, C_{1}, C_{2}, C_{4} \right)
      \right)
  \Biggl\{
  1 + \frac{\alpha _{\rm s}^{\rm{an}}}{\pi}
      \Bigl[ f_{\rm{UV}}\left( x, y, Q^{2}/\mu _{\rm R}^{2}
                                  \right)
\nonumber \\
&&
         + f_{\rm{IR}}\left(x, y, Q^{2}/\mu _{\rm F}^{2}\right)
         + f_{\rm C}(x, y)
            \Bigr]
  \Biggr\}.
\label{eq:NLOpiff}
\end{eqnarray}
In these equations the Sudakov form factor, including evolution, is
given by
\begin{eqnarray}
  S \left(x,y,b,Q, C_{1},C_{2},C_{4}\right)
& \equiv &
   s\left(x, b, Q, C_{1}, C_{2}\right)
 + s\left(\bar{x}, b, Q, C_{1}, C_{2}\right)
 + (x \leftrightarrow y)
\nonumber \\
&&
 - 8 \,\tau \left(C_{1}/b, \mu _{\rm R}\right)
\label{eq:fullSudakov}
\end{eqnarray}
with the ``evolution time''~\cite{SSK00}
\begin{eqnarray}
\everymath{\displaystyle}
  \tau \left( \frac{C_{1}}{b}, \mu \right)
& = &
  \int_{C_{1}^{2}/b^{2}}^{\mu ^{2}} \frac{dk^{2}}{k^{2}} \,
  \frac{\alpha _{\rm s}^{\rm{an}(1)}(k^{2})}{4\pi}
\nonumber \\
& = &
  \frac{1}{\beta _{0}}
  \ln \frac{\ln\left(\mu ^{2}/\Lambda ^{2}\right)}
           {\ln\left(C_{1}^{2}/\left( b\Lambda \right)^{2}
               \right)}
+ \frac{1}{\beta _{0}}
  \left[
  \ln \frac{\mu ^{2}}{\left( C_{1}/b \right)^{2}}
- \ln \frac{\left\vert\mu ^{2} - \Lambda ^{2}\right\vert}
           {\left\vert\frac{C_{1}^{2}}{b^{2}} - \Lambda ^{2}\right\vert}
  \right]
\label{eq:evoltime}
\end{eqnarray}
and the functions $f_{i}$ taken from~\cite{MNP99}. I present
predictions for $F_{\pi}$ in Fig.~\ref{fig:ste_eps2}, adopting the
BLM commensurate-scale method~\cite{BLM83}, and setting $\mu
_{\rm F}= C_{1}/b$ and
$
  \mu _{\rm{BLM}}
=
  \mu _{\rm R} \exp (-5/6),
$
where
$
  \mu _{\rm R}
=
  C_{4} f(x,y) Q
=
  C_{4}\sqrt{xy} Q.
$

%
As one sees, the hard contribution to $F_{\pi}(Q^{2})$ within the AFS
and with a BLM-optimized choice of scales provides a sizeable
fraction of the magnitude of the form factor -- especially at NLO.
No artificial rising at low $Q^{2}$ of the hard form factor appears,
as in conventional approaches, so that this region is dominated by the
Feynman-type contribution~\cite{BRS00}.
Moreover, the self-consistency of perturbation theory has been improved,
as one infers from the saturation behavior of the scaled form factor,
presented on the RHS of Fig.~\ref{fig:ste_eps1}.
Indeed,
$Q^{2}F_{\pi}\left(Q^{2}\right)$
accumulates the bulk of its magnitude below
$b_{\rm cr}\Lambda _{\rm QCD}/C_{1}\leq 0.5$,
i.e., for short transverse distances, where the application of
perturbative QCD is sound.
Even better predictions can be obtained, using a more accurate pion
distribution amplitude, recently derived in~\cite{BMS01} with QCD
sum rules and non-local condensates.

\subsection*{2.2 Power Corrections to Pion Form Factor}
The rationale of {\it global analyticity}, i.e., analyticity as a whole,
implies
\begin{eqnarray}
  \left[ Q^{2}F_{\pi} \left( Q^{2} \right) \right]_{\rm{an}}
& = &
  \int_{0}^{1}\! dx \!
  \int_{0}^{1}\! dy \!
  \Biggl[ \Phi_{\pi}^{\rm out} \!\left( y, Q^{2} \right)
  T_{\rm H}
  \bigl(
        x, y, Q^{2}, \alpha _{s} ( \hat{Q}^{2} )
  \bigr)
  \Phi_{\pi}^{\rm in}\! \left( x, Q^{2} \right)
  \Biggr]_{\rm{an}}
 \nonumber \\
& = &
   A \int_{0}^{1}dx
     \int_{0}^{1}dy \,
  x\, y \bar{x} \bar{y}
  [T_{\rm H}(x, y, Q^{2}, \alpha_{\rm s}( \hat{Q}^{2}))
  ]_{\rm{an}}
\label{eq:ffanaly}
\end{eqnarray}
wherre $A$ is a normalization constant for the pion distribution
amplitude, taken again to be the asymptotic one.
Without the analytization requirement, the pion form factor
is not Borel-summable (see, e.g.,~\cite{Aga96}), but only an
asymptotic series in the coupling constant.
Analytization entails
\begin{equation}
  {\left[
         \alpha _{\rm s}^{n}\left( Q^{2} \right)
  \right]}_{\rm{an}}
\equiv
  \frac{1}{\pi}
  \int_{0}^{\infty}
  \frac{d\xi}{\xi + Q^{2} - i \epsilon}
  \rho ^{(n)}(\xi ),
\label{eq:analycond}
\end{equation}
where the spectral density $\rho ^{(n)}(\xi )$ is the {\it dispersive
conjugate of all powers} $n$ of $\alpha _{\rm s}$.
For the leading-order expression under consideration the spectral
density becomes~\cite{SS97,Rad82,KP82}
\begin{equation}
 \rho \left( Q^{2} \right)
=
  {\rm{Im}} \alpha_{\rm s}\left( - Q^{2} \right)
=
  \frac{\pi}{\beta_{1}}\,
  \frac{1}{\ln ^{2}(Q^{2}/\Lambda^{2}) + \pi^{2}}
\label{eq:specdens}
\end{equation}
Then (\ref{eq:analycond}) reduces to
\begin{equation}
  \left[ \alpha _{\rm s}\left( Q^{2} \right) \right]_{\rm{an}}
=
  \frac{1}{2\pi i}
  \int_{C}^{} \frac{dz}{z - Q^{2} + i \epsilon}
  \alpha _{\rm s}(z),
\label{eq:alphanal}
\end{equation}
where $C$ is a closed contour in the complex $z$-plane with a
branch cut along the negative real axis, assuming exactly
the form of~(\ref{eq:oneloopalpha_an}), as proposed by Shirkov and
Solovtsov~\cite{SS97}.
Recasting the strong coupling in the form
\begin{equation}
  \alpha _{\rm s}(z)
=
  \frac{1}{\beta _{1}}
  \frac{1}{\ln \frac{z}{\Lambda ^{2}}}
=
  \pm \int_{0}^{\infty} d\sigma
      \exp{\left( \mp\sigma \beta _{1} \ln \Lambda ^{2}/z \right)}
\label{eq:expalpha}
\end{equation}
with the plus sign corresponding to the case $|z|/\Lambda ^{2}>1$ and
the minus one to $|z|/\Lambda ^{2}<1$, and inserting it into
(\ref{eq:ffanaly}), we find after some standard manipulations
the Borel transform of the scaled pion form
factor at leading perturbative order~\cite{KS01}:
\begin{equation}
  \left[ Q^{2}F_{\pi}\left( Q^{2} \right) \right]_{\rm{an}}^{(1)}
=
  \int_{0}^{\infty}
  d\sigma
  \exp{\left( - \sigma \beta _{1} \ln Q^{2}/\Lambda ^{2} \right)}
  \tilde{\pi}(\sigma )_{\rm an}^{(1)}.
\label{eq:borelff}
\end{equation}
The Borel image of the form factor reads
\begin{eqnarray}
  \tilde{\pi}(\sigma )_{\rm{an}}^{(1)}
= &&
  16 \pi C_{\rm F} A
  \frac{\sin \left(\pi \beta _{1}\sigma \right)}{\pi}
  \int_{0}^{1} dx
  \int_{0}^{1} dy
  \bar{x}\bar{y}
  \int_{0}^{\infty}
  \frac{d\xi}{\xi + xy}
\nonumber \\
&& \times
  \Biggl[
  \xi ^{-\sigma\beta _{1}}
     \Theta \left( \xi - \frac{\Lambda ^{2}}{Q^{2}} \right)
   + \left( \frac{Q^{2}}{\Lambda ^{2}} \right)^{2\sigma\beta _{1}}
     \xi ^{\sigma\beta _{1}}
     \Theta \left( \frac{\Lambda ^{2}}{Q^{2}} - \xi \right)
  \Biggr].
\label{eq:boreltrans}
\end{eqnarray}
This expression has no IR renormalons in contrast
to approaches that use the conventional one-loop $\alpha _{\rm s}$
parameterization~\cite{Aga96}.

Hence, the integration over the Borel parameter $\sigma$ can be
performed without any ambiguity to arrive at the following result
for the pion form factor
\begin{equation}
  {\left[ Q^{2}F_{\pi}\left( Q^{2} \right) \right]}_{\rm{an}}^{(1)}
=
  16 \pi C_{\rm F} A
  \frac{1}{\beta _{1}}
  \int_{0}^{1} dw\, \phi (w)
  \left[
         \frac{1}{\ln \left( \frac{wQ^{2}}{\Lambda ^{2}} \right)}
        +
         \frac{1}{1-\frac{wQ^{2}}{\Lambda ^{2}}}
  \right].
\label{eq:pifofa1}
\end{equation}
The remaining integration can be carried out analytically to
arrive at an expression derived in~\cite{KS01}. Here I only
display the expression for the physically relevant case
$Q^{2} \gg \Lambda ^{2}$:
\begin{eqnarray}
  {\left[ Q^{2}F_{\pi}\left( Q^{2} \right) \right]}_{\rm an}^{(1)}
& = &
  16 \pi C_{F} A
  \left[
          \frac{1}{4} \alpha _{s}\left( Q^{2} \right)
        + {\cal O}\left( \alpha _{s}^{2} \right)
  \right]
\nonumber \\
&& -
   \frac{1}{\beta _{1}}
   16 \pi C_{F} A
   \frac{\Lambda ^{2}}{Q^{2}}
   \left[
           \frac{1}{2}\ln ^{2}\left( \frac{Q^{2}}{\Lambda ^{2}} \right)
         - 2          \ln     \left( \frac{Q^{2}}{\Lambda ^{2}} - 1 \right)
         - \frac{\pi ^{2}}{3} + 3
   \right]\nonumber \\
&&  + {\cal O}\left( \frac{\Lambda ^{4}}{Q^{4}} \right) \; ,
\label{eq:pifofaQlarge}
\end{eqnarray}
referring for further details to~\cite{KS01}.

\section{Power Corrections to Drell-Yan Process}
As a second example of the AFS, I discuss the derivation of
power corrections to the inclusive Drell-Yan cross-section with the
large scale $Q^{2}$ being here the invariant lepton pair mass.
Citations to previous works and full details of the derivation are
given in~\cite{KS01}.
Consider the logarithmic derivative of the unrenormalized expression
of the eikonalized Drell-Yan cross section, with the notations
of~\cite{KS95}:
\begin{eqnarray}
  \frac{d\ln W_{\rm{DY}}}{d\ln Q^{2}}
& \equiv &
  \Pi ^{(1)}\left( Q^{2} \right)
\nonumber \\
& = &
  4 C_{\rm F} \mu ^{2\epsilon}
  \int_{}^{} \frac{d^{2-2\epsilon}k_{\perp}}{(2\pi )^{2-2\epsilon}}
  \frac{1}{k_{\perp}^{2}} \,
  \alpha _{\rm s}\left( k_{\perp}^{2} \right)
  \left(
        {\rm e}^{-i {\bf k}_{\perp} \cdot {\bf b}} - 1
  \right).
\label{eq:DYcrossec}
\end{eqnarray}
The following important remarks are now in place:
(i) The argument of $\alpha_{\rm s}$ is taken to depend on
$k_{\perp}$ to account for higher-order quantum corrections,
originating from momentum scales larger than this~\cite{KT82}.
(ii) The integral over the transverse momentum is {\it not
well-defined} at very small mass scales owing to the Landau
singularity of the QCD running coupling in that region.
(iii) The evaluation at the edge of phase space is sensitive to
the regularization applied to account for power corrections due
to soft gluon emission transient to nonperturbative QCD.

Imposing analytization as a whole and integrating over transverse
momenta, we obtain
\begin{equation}
  {\left[
         \Pi ^{(1)}\left( Q^{2} \right)
  \right]}_{\rm{an}}
=
  \int_{0}^{\infty} d\sigma \,
  {\rm e}^{- \sigma \beta _{1}
             \ln \left( 4/b^{2}\Lambda ^{2} \right)}
  \tilde{\Pi}_{\rm{an}}^{(1)} (\sigma )
\end{equation}
with a Borel transform given by
\begin{eqnarray}
  \tilde{\Pi}_{\rm{an}}^{(1)}( \sigma )
&& =
  \frac{4 C_{\rm{F}}}{\pi}
  \left(
        \frac{\mu ^{2}b^{2}}{4}
  \right)^{\epsilon}
  \sin \left(
             \pi \sigma \beta _{1}
       \right)
  \int_{0}^{\infty} d\xi \, g( \xi )
  \Biggl[
         \xi ^{- \sigma \beta _{1}}
         \Theta \left(
                       \xi - \frac{b^{2}\Lambda ^{2}}{4}
                \right)
\nonumber \\
&& \;\;\;+
         \left(
               \frac{b^{2}\Lambda ^{2}}{4}
         \right)^{- 2 \sigma \beta _{1}}
         \xi ^{ \sigma \beta _{1}}
         \Theta \left(
                      \frac{b^{2}\Lambda ^{2}}{4} - \xi
                \right)
  \Biggr],
\label{eq:anDYborel}
\end{eqnarray}
where
\begin{equation}
  g( \xi )
=
  \int_{}^{} \frac{d^{2-2\epsilon}q}{( 2\pi )^{2-2\epsilon}}
  \frac{1}{q^{2}}
  \frac{1}{q^{2} + \xi}
  \left(
        {\rm e}^{- 2 i {\bf q} \cdot {\hat{\bf b}}} - 1
  \right).
\label{eq:propagator}
\end{equation}
Combining denominators in Eq.~(\ref{eq:propagator}) and carrying out
the integrations over $\xi$, we then find
\begin{eqnarray}
  {\left[
         \Pi ^{(1)} \left( Q^{2} \right)
  \right]}_{\rm{an}}
&&\!\! =
  \frac{C_{\rm F}}{\pi}
  \left( \mu ^{2}b^{2}\pi
  \right)^{\epsilon}
  \int_{0}^{\infty} d\sigma \,
  {\rm e}^{- \sigma \beta _{1}
  \ln \left(
            4/b^{2}\Lambda ^{2}
      \right)}
  \frac{1}{\Gamma \left( 1 + \sigma \beta _{1} \right)}
\nonumber \\
&& \times
  \Biggl[
         - \frac{1}{\sigma \beta _{1} + \epsilon}
           \Gamma \left( 1 - \sigma \beta _{1} - \epsilon
                  \right)
+     \sum_{n=0}^{\infty}
         \frac{(-1)^{n}}{(n + 1)!}
\nonumber \\
&& \times
         \left(\!
               \frac{b^{2}\Lambda ^{2}}{4}
         \right)^{n + 1 - \sigma \beta _{1} - \epsilon}
         \frac{1}{n + 1 - \sigma \beta _{1} - \epsilon}
  \Biggr]
   -     \frac{C_{\rm F}}{\pi \beta _{1}} \,
         f \left(\! \frac{b^{2}\Lambda ^{2}}{4}
           \!\right)
\label{eq:DYoversigma}
\end{eqnarray}
with $f\left( b^{2}\Lambda ^{2}/4 \right)$ being a complicated
expression, provided in~\cite{KS01} and $\Gamma (x,y)$, denoting the
incomplete Gamma function.
The first term in (\ref{eq:DYoversigma}), viz., the integral
over $\sigma$, diverges for $\sigma \beta _{1}=0$, i.e., for small
values of $\alpha _{\rm s}\left( k_{\perp} \right)$ (or
equivalently for large transverse momenta $k_{\perp}$).
This UV divergence is regulated dimensionally within the
$\overline{MS}$ renormalization scheme adopted here.
Were it not for the terms containing powers of $b \Lambda$,
expression (\ref{eq:DYoversigma}) and that found by Korchemsky and
Sterman~\cite{KS95} (namely, their equation (18)) would be the same.
In our case, however, the imposition of analytization cures all
divergences related to IR renormalons that are generated by the
$\Gamma$-functions whenever $\sigma\beta _{1}$ is an integer different
from zero.

Let us concentrate on the second term in
${\left[
         \Pi ^{(1)} \left( Q^{2} \right)
  \right]}_{\rm{an}}
$
that gives rise to {\it power corrections}. Retaining only
the leading contribution in $b^{2}\Lambda^{2}$, we find
\begin{equation}
  f\left( b^{2}\Lambda ^{2} \right)
=
    - a_{0}
    - a_{1}\frac{b^{2}\Lambda ^{2}}{4}
           \ln\frac{b^{2}\Lambda ^{2}}{4}
    + a_{2}\frac{b^{2}\Lambda ^{2}}{4}
    + {\cal O}\left( b^{4}\Lambda ^{4} \right)
\label{eq:powersfin}
\end{equation}
with the constant coefficients~\cite{KS01}:
$
  a_{0} = 0, \;
  a_{1}\simeq 3.18, \;
  a_{2}\simeq -2.51
$.
Now one can expand the integral in the first term of
${\left[
         \Pi ^{(1)} \left( Q^{2} \right)
  \right]}_{\rm{an}}
$
in powers of $b^{2}\Lambda ^{2}$
and regulate the UV pole at $\sigma\beta _{1}=0$ dimensionally.
For $\sigma\beta _{1}$ an integer, both terms inside the bracket have
poles, but they {\it mutually cancel} so that their sum is
singularity-free and the integral finite.
Retaining terms of order $b^{2}\Lambda ^{2}$, the main
contribution stems from the leading renormalon at
$\sigma\beta _{1}=1$:
\begin{equation}
  {\left[ \Pi ^{(1)}\left( Q^{2} \right) \right]}_{\rm{an}}
=
   {\left[ \Pi ^{(1)}\left( Q^{2} \right) \right]}_{\rm{PT}}
 + {\left[ \Pi ^{(1)}\left( Q^{2} \right) \right]}_{\rm{pow}}
\label{eq:DYfinal}
\end{equation}
with the perturbative part being defined by
\begin{equation}
   {\left[ \Pi ^{(1)}\left( Q^{2} \right) \right]}_{\rm{PT}}
=
   \frac{C_{\rm F}}{\pi\beta _{1}}
   \ln \frac{\ln \left( C/b^{2}\Lambda ^{2} \right)}
            {\ln \left( Q^{2}/\Lambda ^{2} \right)},
\label{eq:DYfinpt}
\end{equation}
a result coinciding with that obtained in~\cite{KS95}.
Power corrections in the impact parameter $b$ are encoded in the
second contribution ($b^{2}\Lambda ^{2}\ll 1$):
\begin{equation}
  {\left[ \Pi ^{(1)}\left( Q^{2} \right) \right]}_{\rm{pow}}
=
    S_{0} + b^{2} S_{2}\left( b^{2}\Lambda ^{2} \right)
  + {\cal O}\left( b^{4}\Lambda ^{4}
\right),
\label{eq:DYpow}
\end{equation}
where
\begin{equation}
  S_{0}
=
  \frac{C_{\rm F}}{\pi\beta _{1}} a_{0} = 0
\label{eq:powconst}
\end{equation}
and
\begin{equation}
  S_{2}\left( b^{2}\Lambda ^{2} \right)
 =
  \frac{C_{F}}{4\pi\beta _{1}} \Lambda ^{2}
    a_{1}\ln \frac{b^{2}\Lambda ^{2}}{4}
  - a_{2}
\label{eq:leadpow}
\end{equation}

The DY cross-section $W_{\rm DY}$, comprising the leading logarithmic
perturbative contribution (Sudakov exponent $S_{\rm{PT}}$) and
the first power correction (in $b^{2}\Lambda ^{2}$) reads (with the
$Q$-dependence arising due to collinear interactions)
\begin{equation}
  W_{\rm DY}(b, Q)
=
  \exp \left[
             - S_{\rm PT}(b, Q) - b^{2} S_{2}(b,Q) + \ldots
       \right] \; ,
\label{eq:DYCSfinal}
\end{equation}
where
\begin{equation}
  S_{2}(b,Q)
\sim
   S_{2}\left( b^{2}\Lambda ^{2} \right)\ln Q + \rm{const}.
\label{eq:s_2}
\end{equation}
The Sudakov factor, representing the perturbative tail of the
hadronic wave function, {\it suppresses} constituent
configurations which involve large impact space separations,
while the exponentiated power corrections in $b^{2}$, being of
nonperturbative origin, {\it provide enhancement} of such
configurations (since
$S_{2}\left(b^{2}\Lambda ^{2}\right)$ is always negative).
Hence, the net result is less suppression of the DY cross-section
and also enhancement of the pion wave function in $b$ space with
the endpoint region $b\Lambda \sim 1$ being less enhanced relative to
small $b$ transverse distances.

\section{Conclusions}
I have presented a theoretical framework, based on {\it analytization}
that enables the calculation of perturbative gluonic corrections
(Sudakov form factor), as well as power-behaved ones that are linked to
nonperturbative effects in QCD. Moreover, one can calculate the
absolute normalization of the power corrections to hadronic observables
{\it systematically} without any renormalon ambiguity from the
outset.

\vspace{0.3cm}
\acknowledgments
I wish to thank the organizers for the warm hospitality extended
to me during this stimulating meeting and the Deutsche
Forschungsgemeinschaft for a travel grant. It is a pleasure to thank
A.P. Bakulev, A.I. Karanikas, and W. Schroers for collaboration and
K. Passek, M. Prasza{\l}owicz, D.V. Shirkov, and I.L. Solovtsov for
discussions.

%


\begin{thebibliography}{8.}
\addcontentsline{toc}{section}{References}
\bibitem{CS81} J.C. Collins, D.E. Soper:
               Nucl. Phys. B {\bf 193}, 381 (1981);
               ibid. B {\bf 197}, 446 (1982)

\bibitem{KS95} G.P. Korchemsky, G. Sterman:
               Nucl. Phys. B {\bf 437}, 415 (1995)
               [hep-ph/9411211]

\bibitem{CSS85} J.C. Collins, D.E. Soper, G. Sterman:
                Nucl. Phys. B {\bf 250}, 199 (1985)

\bibitem{KS01} A.I. Karanikas, N.G. Stefanis:
               Phys. Lett. B {\bf 504}, 225 (2001)
               [hep-ph/0101031]

\bibitem{Ste99} N.G. Stefanis:
                Eur. Phys. J.direct C {\bf 7}, 1 (1999)
                [hep-ph/9911375]

\bibitem{Shi01} D.V. Shirkov:
                Eur. Phys. J. C {\bf 22}, 331 (2001)
                [hep-ph/0107282];
                Theor. Math. Phys. {\bf 127}, 409 (2001)
                [hep-ph/0012283]

\bibitem{SS99} D.V. Shirkov and I.L. Solovtsov,
               Theor. Math. Phys. 120 (1999) 1210
               [Teor. Mat. Fiz. 120 (1999) 482]
               [hep-ph/9909305].

\bibitem{SS97} D.V. Shirkov, I. L. Solovtsov:
               Phys. Rev. Lett. {\bf 79}, 1209 (1997)
               [hep-ph9704333]

\bibitem{SSK00} N.G. Stefanis, W. Schroers, H.-Ch. Kim:
                Eur. Phys. J. C {\bf 18}, 137 (2000)
                [hep-ph/0005218]

\bibitem{LS92} H.-n. Li, G. Sterman:
               Nucl. Phys. B {\bf 381}, 129 (1992)

\bibitem{BHL83} S.J. Brodsky, T. Huang, G.P. Lepage. In:
                {\it Banff Summer Institute on Particles and Fields (1981)},
                ed. by A.Z. Capri and A.N. Kamal
                (Plenum Press, New York 1983) pp. 143-199.

\bibitem{PR01} M. Prasza{\l}owicz, A. Rostworowski:
               Phys. Rev. D {\bf 64}, 074003
               [hep-ph/0105188];
               [hep-ph/0111196]

\bibitem{KR87} G.P. Korchemsky, A.V. Radyushkin:
               Nucl. Phys. B {\bf 283}, 342 (1987)

\bibitem{Kor89} G.P. Korchemsky:
                Phys. Lett. B {\bf 217}, 330 (1989);
                ibid. B {\bf 220}, 629 (1989)

\bibitem{GKKS97} G. Gellas, A.I. Karanikas, C.N. Ktorides,
                 N.G. Stefanis:
                 Phys. Lett. B {\bf 412}, 95 (1997)
                 [hep-ph/9707392];
                 A.I. Karanikas, C.N. Ktorides, N.G. Stefanis,
                 S.M.H. Wong:
                 Phys. Lett. B {\bf 455}, 291 (1999)
                 [hep-ph/9812335]

\bibitem{KT82} J. Kodaira, L. Trentadue:
               Phys. Lett. B {\bf 112}, 66 (1982)

\bibitem{MNP99} B. Meli\'c, B. Ni\v zi\'c, K. Passek:
                Phys. Rev. D {\bf 60}, 074004 (1999)
                [hep-ph/9802204]

\bibitem{BLM83} S.J. Brodsky, G.P. Lepage, P.B. Mackenzie:
                Phys. Rev. D {\bf 28}, 228 (1982)

\bibitem{Bro73} C.N. Brown et al.:
                Phys. Rev. D {\bf 8}, 92 (1973)

\bibitem{Beb76} C.J. Bebek et al.:
                Phys. Rev. D {\bf 13}, 25 (1976) 25;
                ibid. D {\bf 17}, 1693 (1978)

\bibitem{BRS00} A.V. Radyushkin:
                Few Body Syst. Suppl. {\bf 11}, 57 (1999)
                [hep-ph/9811225];
                A.P. Bakulev, A.V. Radyushkin, N.G. Stefanis:
                Phys. Rev. D {\bf 62}, 113001 (2000)
                [hep-ph/0005085]

\bibitem{BMS01} A.P. Bakulev, S.V. Mikhailov, N.G. Stefanis:
               Phys. Lett. {\bf B508}, 279 (2001)
               [hep-ph/0103119];
               hep-ph/0104290

\bibitem{Aga96} S. Agaev,
                hep-ph/9611215;
                Mod. Phys. Lett. A {\bf 13}, 2637 (1998)
                [hep-ph/9805278];
                Eur. Phys. J. C {\bf 1}, 321 (1998)
                [hep-ph/9611283]

\bibitem{Rad82} A.V. Radyushkin:
                JINR preprint E2-82-159, JINR Rapid Communications
                4[78]-96, 9 (1982)
                [hep-ph/9907228]

\bibitem{KP82} N.V. Krasnikov, A.A. Pivovarov:
               Phys. Lett. B {\bf 116}, 168 (1982)
\end{thebibliography}
\end{document}